\documentclass[graphicx]{emulateapj}
\usepackage{apjfonts}

\slugcomment{to appear in ApJ}
\shortauthors{Steeghs et al.}
\shorttitle{The component masses in WZ Sge}

\begin{document}

\title{Dynamical constraints on the component masses of the cataclysmic variable WZ Sge}

\author{Danny Steeghs,\altaffilmark{1,4} 
Steve B. Howell,\altaffilmark{2}, 
Christian Knigge,\altaffilmark{3} 
Boris T. G\"ansicke,\altaffilmark{4} 
Edward M.Sion,\altaffilmark{5}, William F.Welsh,\altaffilmark{6} }

\altaffiltext{1}{Harvard-Smithsonian Center for Astrophysics, 60 Garden Street, Cambridge, MA 02138 
{\it dsteeghs@cfa.harvard.edu}}
\altaffiltext{2}{WIYN Observatory and National Optical Astronomy Observatory,
950 N. Cherry Ave, Tucson, AZ 85719 {\it howell@noao.edu}}
\altaffiltext{3}{Dept. of Physics and Astronomy, University of Southampton,
Highfield, Southamption, SO17 1BJ, UK {\it Christian@astro.soton.ac.uk}}
\altaffiltext{4}{Department of Physics, University of Warwick, Coventry
CV4 9BU, UK {\it Boris.Gaensicke@warwick.ac.uk}}
\altaffiltext{5}{Department of Astronomy and Astrophysics, Villanova University, 
800 Lancaster Ave. Villanova, PA, 19085 {\it emsion@ast.villanova.edu}}
\altaffiltext{6}{Department of Astronomy, San Diego State University, San Diego, CA
{\it wfw@sciences.sdsu.edu}}

\begin{abstract}

We present phase-resolved spectroscopy of the short period
cataclysmic variable WZ Sge obtained with the {\it Hubble
Space Telescope}. We were able to resolve the orbital
motion of a number of absorption lines that likely probe
the environment near the accreting white dwarf. The radial
velocities derived from simultaneous fits to 13 absorption
lines indicate an orbital velocity semi-amplitude of
$K_{UV}=47\pm3$ km/s. However, we find that the phase zero
is offset from the white dwarf ephemeris by +0.1.
Our offset and velocity amplitude are very similar to
constraints derived from optical emission lines from the
quiescent accretion disk, despite the fact that we are
probing material much closer to the primary. If we
associate the UV amplitude with $K_1$, our dynamical
constraints together with the $K_2$ estimates from Steeghs
et al. (2001) and the known binary inclination of $i=77\pm2$ imply 0.88$<M_1<$1.53$M_{\odot}$,
$0.078 < M_2 < 0.13 M_{\odot}$ and
$0.075<q=M_2/M_1<0.101$. If we interpret the mean velocity
of the UV lines ($-16\pm4$ km/s) as being due to the
gravitational red-shift caused in the high-$g$ environment
near the white dwarf, we find $v_{grav}=56\pm5$km/s which
provides an independent estimate on the mass of the primary
of $M_1=0.85\pm0.04M_{\odot}$ when coupled with a mass-radius relation. Our primary mass estimates
are in excellent agreement and are also self-consistent
with spectrophotometric fits to the UV fluxes despite the
observed phase offset. It is at this point unclear what
causes the observed phase-offset in the UV spectra and by
how much it distorts the radial velocity signature from the
underlying white dwarf.

\end{abstract}


\section{Introduction}

WZ Sge is a crucial and well-studied member of the cataclysmic variables (CVs), harboring a white dwarf primary accreting via Roche lobe overflow from a low mass companion. Its long outburst recurrence time ($\sim$30 yrs), large outburst amplitude ($\sim$7 mags), short orbital period (82 mins) and low mass accretion rate have long been associated with the picture of an evolved CV close to its period minimum (see Kato et al. 2001 and Patterson et al. 2002 for reviews). 
%
%
With a distance of only 43.5 pc (Thorstensen 2003, Harrison et al. 2004), it is the closest known cataclysmic variable and reaches V$\sim$8 near its outburst peak. Most of its life, WZ Sge sits near V$\sim$15 corresponding to a rather modest M$_V\sim$12. Its proximity makes WZ Sge a rewarding object of study even if its characteristics among the growing family of short orbital period CVs are no longer that extreme or unusual. 
The rare, large amplitude outbursts and the absence of regular dwarf nova outbursts have now been seen in a small group of WZ Sge-like systems (Patterson et al. 2005a; Templeton et al. 2006), making these binaries particularly interesting from the point of view of understanding the disk instability mechanism that is thought to drive outbursts in CVs, X-ray transients and FU Orionis stars.
%

Attempts at determining the component masses and thereby verify the evolutionary status of WZ Sge have long been frustrated without the detection of direct spectroscopic radial velocity signatures tracing either of the two stars.
Published mass values for the white dwarf have ranged from $0.45M_{\odot}$
(Smak 1993) to $\sim1.1M_{\odot}$ (Gilliland, Suntzeff \&
Kemper 1986).  Spruit \& Rutten (1998) used the broad
disk emission lines as well as the bright spot dynamics and find primary masses from $0.9M_{\odot}$ to as high as $1.2M_{\odot}$ depending on the assumptions concerning the nature of the hotspot tail.
Indirect estimates using the dynamics of the emission lines may not reflect the true dynamical motion of the white dwarf itself as was illustrated by significant phase offsets between the motion in the disk lines and that expected for the primary.
The donor star also proved elusive despite the low mass accretion rate during quiescence implying an intrinsically faint and low mass object (e.g. Howell et al. 2004). However, during the 2001 outburst activity, Steeghs et al. (2001) were able to detect sharp emission features originating from the irradiated hemisphere of the donor, thus providing the first direct trace of its orbital dynamics corresponding to mass limits of  $M_2<0.11M_{\odot}$ and $M_1 > 0.7M_{\odot}$.

If we can determine the mass of the white dwarf star directly by measuring the
radial velocities of its photospheric absorption lines, then WZ Sge offers
parameter solutions of a double-lined spectroscopic binary.  As part of
our monitoring of the cooling of the white dwarf in response to the July
2001 superoutburst (Sion et al. 2003; Long et al. 2004), five HST orbits were dedicated to phase-resolved
spectroscopy to attempt a mass determination of the WZ Sge white dwarf.
In this Letter we report the results of this spectroscopy.

\section{Observations and Data Reduction}

WZ Sge was observed on five consecutive orbits with the {\it Hubble Space
Telescope} during 2004 July 10-11. We used STIS and the E140M echelle
grating with the 0.2" x 0.2" aperture. This setup yielded a spectral
resolution of $\sim$90,000 and covered the range of 1150-1710\AA. Each
orbit consisted of approximately 2800 total seconds of on-source
observation.
We refer to Godon et al. (2006) for a discussion of the observations as well as the time-averaged properties of WZ Sge derived from these data.
For the purposes of our phase-resolved analysis, we divided each of the five orbits of STIS
time-tagged spectroscopy into four summed datasets giving us 20 individual
spectra to work with. Orbit one was slightly shorter than the other four,
and we divided the spectra into four 575 sec bins. The
final four orbits were all equal in length and we divided them into 16
spectra of 712 sec each. 
These spectra provided full-phase coverage across the 82 min binary orbit. All spectra were transfered into a helio-centric frame by applying both velocity and time-system corrections. The orbital phases were then calculated using the photometric bright spot ephemeris of Patterson et al. (1998b) shifted by -0.046 in orbital phase in order to make phase zero coincident with conjunction of the donor star (Steeghs et al. 2001). The uncertainty on this ephemeris is limited by the phase zero correction which is good to 0.03 binary orbits.

The absolute wavelength calibration of STIS is expected to be good to less than a pixel provided the target is well centered in the slit. The slit acquisition precision should also be good enough for a relatively bright source such as WZ Sge and cause at most a 1 pixel shift in the zero-point. For our E140M echelle data this amounts to a potential zero-point shift of up to 3km/s. 
We verified this by fitting the position of the geocoronal Ly-$\alpha$ emission and found that its position was only 1 km/s off from the nominal velocity of the Earth at the time of the observations. We therefore consider the absolute wavelength scale to be accurate to better than 3 km/s.

The average spectrum of WZ Sge as derived from our 20 STIS spectra is 
presented in Figure 1. Here we have removed the orbital velocities of the white dwarf absorption lines from the individual spectra before averaging using the radial velocity amplitudes derived in this paper. The spectrum shows a rich absorption line spectrum plus broad emission features from the accretion flow (see also Godon et al. 2006 for line identifications).

\begin{figure}[t]
\includegraphics[height=8.5cm, angle=-90]{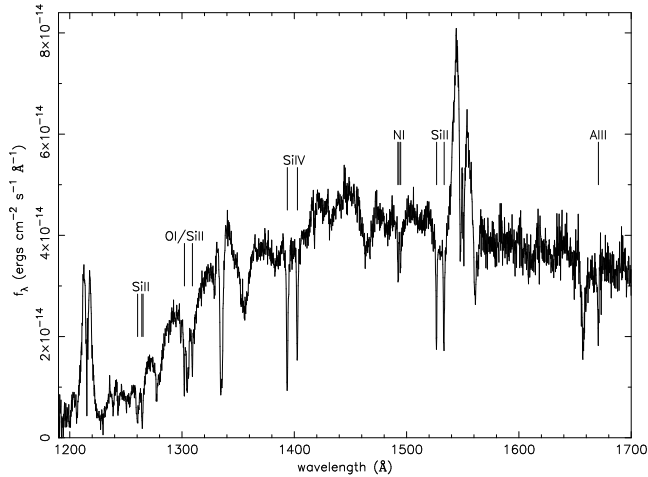}
\caption{Average HST spectrum of WZ Sge obtained in July 2004. The orbital velocities of the narrow absorption lines were removed from each individual spectrum before averaging using the radial velocity amplitude derived in this Letter. The thirteen spectral lines that were used in our final radial velocity fits are marked by vertical ticks.  \label{fig1}}
\end{figure}

\section{Analysis}

In order to search for radial velocity signatures in the rich absorption line spectrum that is associated with the primary white dwarf, we first re-binned the data to a constant velocity-scale wavelength grid, and subsequently normalized the continuum. The continuum shape was derived from a 2nd order spline-fit to line free regions of the spectrum. To boost the signal to noise further, the normalized spectra were then phase-folded into variance weighted bins to provide 10 phase bins across the binary orbit.
%
%

We investigated the time dependent properties of a range of spectral lines using a variety of techniques, including by-eye centroid estimates, formal Gaussian fits to individual lines and cross-correlation techniques.
We found significant radial velocity changes between the individual spectra with a semi-amplitude of approximately 50 km/s and close in phase to what is expected for the white dwarf. The formal Gaussian fits to the line profiles provided the most reliable and stable velocities, and we thus choose this as our preferred method.
A large number of strong absorption lines are present in the spectra, although only those lines with unblended line cores were suitable for Gaussian fitting. For this reason, we are not including the strong carbon blends in our analysis.
When performing the Gaussian fits, the width was fixed to a value estimated from the average spectrum leaving the line center and strength as the two free parameters. This fixed width value was then revised by shifting out the fitted velocities and re-fitting for the width using the average derived after shifting the individual bins.
Starting with the strongest lines, we found that individual spectral lines displayed significant and periodic radial velocities as a function of the binary orbit. Even more encouraging, the phasing and amplitudes derived from different lines were in agreement. It thus appeared that these spectral lines are all tracking a common radial velocity curve. Since these features are expected to be formed close to the photosphere of the white dwarf, they appear to be promising tracers of its orbital motion.

In order to improve the signal to noise of our radial velocities, we grouped lines together in order to perform simultaneous fits to multiple lines. The free parameters in these fits involving more than one line were the strength of each line as well as a {\it single} common velocity offset for the set of lines fitted, relative to their respective rest wavelengths.
In all our Gaussian fits, formal fitting errors on the fit parameters were calculated based on the variances of each individual data point used in the fit. 
We tried different groups of lines in order to check how stable the derived velocities were to the specific sets of lines used and found that all the lines seem to share the same orbital kinematics.
This led us to select a set of 13 absorption lines in the spectrum of WZ Sge that were well resolved and strong enough to be fitted with Gaussians in the individual spectra. We list the wavelengths of the 13 features used in these joint fits as well as the widths of the employed Gaussians in Table 1 and the resulting radial velocity measurements in Table 2.
To further test the reliability of our fits, we also re-binned the data at various levels to test the robustness of our velocity values to various levels of binning. We found that the derived radial velocity variations were only weakly sensitive to the binning of the data and the variations were within the statistical uncertainties associated with the fits. Some level of binning is preferred to avoid fitting to very low S/N spectra and for the actual velocity values reported in this study, we worked with individual spectra binned to a constant velocity dispersion of 20 km/s/pixel.
In Figure 2, we present the derived radial velocities as a function of the orbital phase of the binary, together with the best fit sinusoidal radial velocity curve. These fits were with fixed period and the zero-phase, the amplitude and the mean velocity as free parameters. For comparison, we have plotted the results from fitting to just two strong spectral lines as well as our final fit that involves the 13 spectral lines listed in Table 1. 
All other trial fits to several groups of lines show the same kinematics, with amplitudes varying between 44-51 km/s and 0.10-0.14 phase offsets relative to the binary ephemeris. It thus appears that despite the fact that the origin of all the observed lines is not fully understood (Long et al. 2004), they do share the same orbital kinematics. 
Of particular interest are the strong Si {\sc IV} doublet lines near 1400\AA~(Fig 2) that track the Si {\sc II} and other lines perfectly even though Si {\sc IV} requires temperatures in excess of 25,000K while the photospheric temperature of the WD is close to 15,000K. 
However, zero phase is consistently and significantly offset from what is expected for material co-moving with the white dwarf's center of mass by $\sim0.1$. Our formal phase offset uncertainty is only 0.02 and the uncertainty in the orbital ephemeris is less than 0.03. 
We comment that the error bars plotted in Fig.2 reflect the formal fitting errors for our joint Gaussian fits. When comparing different group of lines, some systematic residuals from a sinusoidal radial velocity are observed. However, these additional systematic effects are modest as we typically achieve goodness of fit values close to $\chi^2_\nu=1$. For our final joint fit to 13 lines for example, we find a formal $\chi^2_\nu=1.4$.

We conclude that we have detected sinusoidal velocity shifts in a large number of absorption lines that are expected to be formed very close to the primary. They should be the most reliable tracers we currently have for the orbital dynamics of the primary, but the $\sim0.1$ phase offset suggests that these lines appear to be contaminated by line of sight components other than the underlying white dwarf photosphere.

From our joint fit, we derive a mean velocity of  $-16 \pm 3$ km/s, a radial velocity amplitude of $47 \pm 3$ km/s and a phase zero offset of $+0.12 \pm 0.02$ with the errors corresponding to formal $1\sigma$ uncertainties on the fit parameters. 
Given that the absolute wavelength scale is accurate to 3 km/s, we added this additional uncertainty in quadrature to the formal least-squares error of the mean velocity which brings its final uncertainty to 4 km/s.

\begin{table}
\begin{center}
\caption{Lines included in dynamical fits}
\begin{tabular}{lcc}
\tableline\tableline
Ion & rest wavelength & FWHM (km/s) \\
\tableline
Si II & 1260.4221 & 500\\
Si II & 1264.7377 & 500\\
Si II & 1265.0020 & 500\\
O I & 1302.1680 & 175\\
Si II & 1309.2760 & 175\\
Si IV & 1393.7546 & 250\\
Si IV & 1402.7697 & 250\\
N II & 1492.6254 & 250\\
N II & 1492.8200 & 250\\
N II & 1494.6751 & 250\\
Si II & 1526.7066 & 250\\
Si II & 1533.4310 & 250\\
AlII & 1670.7870 & 250\\
\tableline\tableline
\
\end{tabular}
\end{center}

\end{table}
\begin{table}
\begin{center}
\caption{Radial velocities}
\begin{tabular}{lcc}
\tableline\tableline
Phase & velocity (km/s) & $\sigma$ \\
\tableline
0.010  & -00.09   & 12.1 \\
0.140  & -23.70   & 13.5 \\
0.190  & -23.32   & 10.7 \\
0.258  & -58.39   & 9.4 \\
0.370  & -60.61   & 9.2 \\
0.507  & -58.27   & 7.3 \\
0.601  & -42.26   & 9.9 \\
0.704  &  10.09   & 6.3 \\
0.819  &  29.29   & 12.8 \\
0.865  &  11.69   & 9.8 \\
\tableline\tableline
\
\end{tabular}
\end{center}

\end{table}

\begin{figure}[t]
\includegraphics[height=8.5cm,angle=-90]{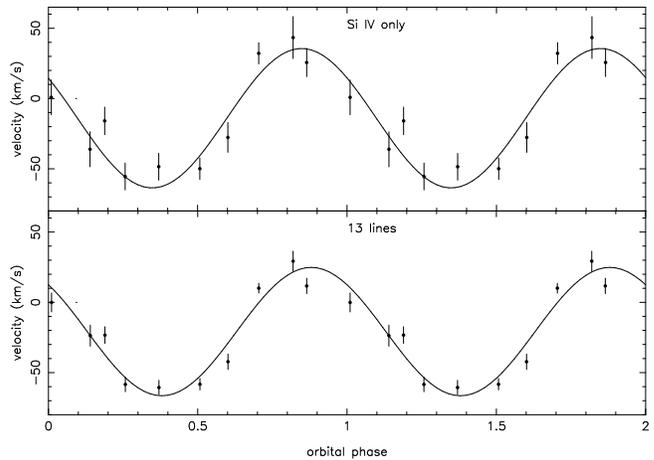}
\caption{Radial velocity measurements of the photospheric absorption features from the accreting white dwarf. Top panel are the velocities derived from fitting to the two strong Si IV features only whereas the bottom panel shows our final velocities derived from simultaneous fits to thirteen spectral lines. Least-squares sine fits are overplotted. \label{fig2}}
\end{figure}

\section{Dynamical constraints on the component masses}

We were able to resolve the orbital motion of the UV absorption line system in WZ Sge using our high-resolution STIS time-series data. Unfortunately, the radial velocity curve displays a significant phase offset with the radial velocity curve expected for the primary and we must therefore be careful with associating our derived semi-amplitude with the orbital velocity of the white dwarf: $K_{UV}=47\pm3$ km/s. 
The bright-spot eclipse has always served as a reliable clock for the system (Patterson et al. 1998). It is difficult to see how the accurate bright spot ephemeris combined with the correction to true orbital phase zero as provided by the detection of the mass donor star (Steeghs et al. 2001) could lead to a significant error in the calculation of the orbital phases and we estimated that our shifted ephemeris should be good to 0.03 cycles. Also while the STIS absolute timing accuracy can drift somewhat, it is not expected to approach the 8 minutes of clock error we would need to undo the 0.1 phase shift we observe. 
Interestingly enough, our amplitude {\it and} phase offset is consistent with previous $K_1$ estimates reported by Gilliland et al. (1986), Spruit \& Rutten (1998), Mason et al. (2000) and Steeghs et al. (2001). 
Those estimates relied on optical emission from the quiescent accretion disc and is based on the kinematics of the disk gas several to several tens of white dwarf radii out. The emission region we are studying in the UV on the other hand originates very close to the white dwarf surface.
It appears that despite our use of UV lines formed much closer to the primary, a very similar radial velocity behavior is observed. 

In this and the next section, we consider the implications our UV constraints would have on the system parameters of WZ Sge. We will revisit the reliability of these constraints given the observed phase offset in the discussion section.

If we consider our amplitude estimate from the UV spectroscopy as an estimate for $K_1$ and combine it with the conservative donor star velocity estimates ($493 < K_2 < 585$ km/s) of Steeghs et al. (2001), we derive mass limits for the primary ranging between $M_1 \sin^3{i} = 0.83M_{\odot} $ and $M_1 \sin^3{i} = 1.38M_{\odot}$. 
Similarly, it would constrain the donor star to the range $ 0.074M_{\odot}  < M_2 \sin^3{i} < 0.12 M_{\odot}$.  Only our dynamical constraints on $K_1$ and $K_2$ have so far entered these mass estimates.
Additional constraints are provided by the prominent bright spot eclipses, even though the primary itself is never eclipsed. This limits the binary inclination range considerably since the lack of a white dwarf eclipse sets an upper limit to the allowed inclination, whereas the prominent bright spot eclipse indicates the inclination needs to be large enough to obscure the outer disc.
Several inclination estimates have been reported in the literature. Skidmore et al. (2002) modeled the infrared eclipse lightcurve in quiescence and finds $i=75.9$, Spruit \& Rutten (1998) quote $i=77\pm2$.
For $i=75-79$, our formal mass limits correspond to $0.88M_{\odot}<M_1<1.53M_{\odot} $ and $0.078M_{\odot} <M_2<0.13M_{\odot} $.

\section{M$_1$ from the Gravitational Redshift}

Apart from the radial velocity amplitude, the second insight provided by the observed absorption line velocities is the recovered mean velocity of $-16 \pm 4$ km/s. 
Previous studies of the optical emission lines using both the broad disk emission (Spruit \& Rutten 1998) as well as the narrow emission component from the mass donor star (Steeghs et al. 2001) indicate a systemic binary velocity of $\gamma = -72 \pm 2$ km/s relative to a heliocentric velocity frame. The UV absorption  lines are thus shifted by $+56 \pm 5$ km/s. This would be expected for lines formed near the primary due to the gravitational redshift introduced in the high-$g$ environment of the white dwarf (Eddington 1924; Greenstein \& Trimble 1967; Sion et al. 1994).
The gravitational redshift is proportional to $M/R$ and can thus provide a completely independent determination of the primary mass given a mass-radius relation. Assuming that the lines are formed at or close to the surface of the primary and using Eggleton's zero-temperature mass radius relation as quoted in Verbunt \& Rappaport (1988) implies $M_1=0.84\pm 0.04 M_{\odot}$. 
The above mass radius relation is expected to be relevant for the WD masses considered here, but does not consider the possible effects of the internal chemical composition of the white dwarf and the presence of substantial surface layers of H and He.
We have no direct constraints on the internal composition of the white dwarf in WZ Sge, but stellar and binary evolution models tell us that a CO dominated WD is expected with relatively thin and low mass H/He shells. Regular nova eruptions keep the shells thin and of low mass fraction and for the purpose of this paper we do not attempt to constrain such composition issues using our redshift measurement. 
In Figure 3, we illustrate the possible effects of WD composition and temperature on the expected mass-radius relation, and thus the inferred mass.  We used the models of Panei, Althaus \& Benvenuto (2000) to plot mass-radius curves for various compositions. These provide models for He,C and O dominated white dwarfs with possible H and He layers for a range of temperatures. Given that WZ Sge's white dwarf was close to 15,000K at the time of our observations (Godon et al. 2006), we selected relevant models at 15,000K and plot a family of model curves in Fig.3.
It can thus be seen that temperature and composition effects are relatively modest in comparison to our formal error on $v_{grav}$. If we take C or O dominated models at 15,000K and allow for the presence of modest H and He layers the inferred mass is pushed up slightly, but $M_1=0.85\pm 0.04M_{\odot}$ encompasses the relevant models within $v_{grav}$=56$\pm$5 km/s.

We comment that the mean red-shift observed reflects the $M/R$ at the radius where the lines are formed, which may not necessarily be the canonical white dwarf surface. If the lines are instead formed away from the surface at larger radii we would underestimate the white dwarf mass since we overestimate its radius.
There are therefore potential systematic effect at play that will affect our inferred mass from the gravitational red-shift which are not reflected in the 5\% formal error quoted. The key purpose in this paper is to provide a complementary estimate for the mass of the white dwarf, that can be compared to the masses inferred from the orbital velocities.
Our primary mass/radius estimates based on the gravitational redshift interpretation of the mean velocity are in excellent agreement with recent radius estimates derived from spectrophotometric fits to FUSE and HST spectra. Given our knowledge of the distance to WZ Sge, such fits directly constrain the size of the emitting region and achieve self-consistent solutions between the required log $g$ and the radius for a white dwarf mass close to $0.9M_{\odot}$ (Long et al. 2003; Long et al. 2004; Godon et al. 2006).

\begin{figure}[t]
\includegraphics[height=8.5cm, angle=-90]{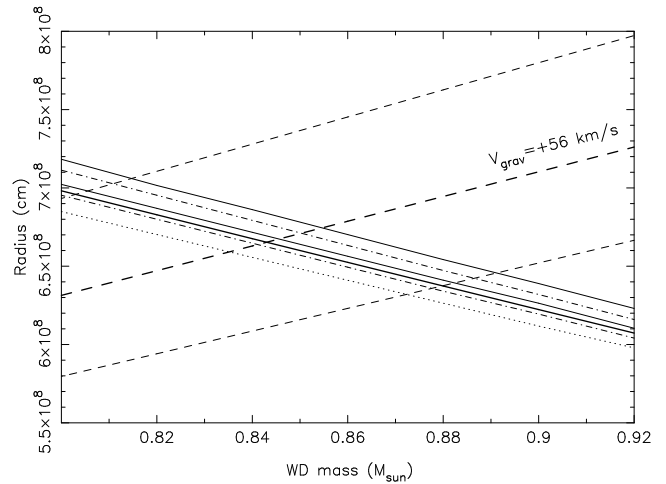}
\caption{Mass-radius constraints for the primary based on our gravitational redshift measurement. The labeled thick dashed line represent the $V_{grav}=56$km/s curve with additional dashed curves for the $\pm1\sigma$ constraints. The zero-temperature Hamada-Salpeter  mass-radius relation is shown as the solid curve. Additional models from Panei et al. (2000) show the dependence on WD composition and temperature. Thin solid curves are for carbon composition comparing a He layer only (lower curve) with a He and a H layer at 15,000
K. The dot-dashed curves are the same for oxygen composition. Finally, the dotted curve is a O model with a He layer at 5,000K.  \label{fig3}}
\end{figure}

\section{Discussion}

A robust determination of the white dwarf radial velocity amplitude, $K_1$, is the key to reliable system parameter estimates for WZ Sge. WZ Sge has always played an important role in our understanding of CV evolution, but despite its bright outbursts and proximity there is still considerable uncertainty about the mass and nature of its companion star. 
Our HST observations did reveal diagnostic UV absorption lines that display stable orbital kinematics, but suffer from the same phase offset as previous estimates using the accretion disk emission lines in the optical. It is at this point unclear what causes the observed phase-offset in the UV spectra and by how much it distorts the radial velocity signature from the underlying white dwarf. The observed Si {\sc IV} lines also demand formation temperatures well above that of the underlying photoshpere. We must therefore be cautious with our derived system parameter estimates.

The observed radial velocity amplitude implies a primary mass of at least $0.83 M_{\odot}$ and a binary mass ratio $0.075<q<0.10$ if we ignore the phase offset and assume we are indeed measuring $K_1$. In addition to our estimate for $K_1$, we derived an independent  estimate for the mass of the primary if one interprets the mean velocity as corresponding to the gravitational redshift near the white dwarf surface. This implied $M_1=0.85\pm 0.04 M_{\odot}$.  If we wish to make the system parameters consistent with both our $K_1=47 \pm 3$ as well as $M_1=0.85 \pm 0.04 M_{\odot}$ from the redshift, we require $q=0.092 \pm 0.008$, $M_2=0.078 \pm 0.06 M_{\odot}$ and $K_2=510 \pm 15$ km/s. 
Despite the phase offset, there is excellent agreement between our dynamical mass estimates, the gravitational redshift inferred mass as well as the log $g$ and emitting radius constraints from fits to the UV fluxes given the known distance to WZ Sge.
A white dwarf mass of $0.85M_{\odot}$ holds an important implication
for the accretion heating of the white dwarf. Quasi-static evolutionary
models with compressional heating, boundary layer irradiation and time
variable accretion by Godon et al. (2006) require additional heating mechanisms (e.g. ongoing accretion after the outburst,
boundary layer irradiation among others) to bring the evolutionary
models in agreement with the observed cooling for a primary mass of $0.8-0.9M_{\odot}$. 

Although the various published estimates for the white dwarf mass in WZ Sge tended to converge near $0.9M_{\odot}$, in line with our estimates, it has long been thought that the mass donor star was much less massive and accordingly that the mass ratio was very small. Indeed, WZ Sge was considered one of the prime candidates for being an evolved CV that has whittled down its donor star and has evolved through its orbital period minimum (Ciardi et al. 1998; Patterson 1998a). Depending on the precise value for the period minimum, a matter that has not yet been settled, a post-period bounce system at the period of WZ Sge would have a donor mass of $0.04-0.06M_{\odot}$ (Kolb \& Baraffe 1998; Howell 2001). 
The first solid example of such a system has only recently been discovered in the sample of SDSS CVs (Littlefair et al. 2006). We now have a growing sample of CVs accreting at low rates, with periods near the period minimum. Several of these display rare, large amplitude dwarf nova outbursts in contrast with the more prolific outburst behaviour at longer orbital periods and higher mass transfer rates. There is still a debate ongoing whether the long outburst recurrence times of these systems require an unusually low viscosity during the quiescent phases (Smak 1993), or if the inner disk can be stabilised or depleted (Hameury et al. 1997; Matthews et al. 2007).
Given the intrinsic faintness of the short period systems, and the fact that a significant sample of such systems has only been inventarised very recently, their long-term variability has not been that well characterised. But it is clear that WZ Sge is part of a class of evolved CVs and that system parameter studies of these systems can provide important clues concerning our understanding of CV evolution and the nature of the disk instability process.

The mass ratio implied by our constraints is rather high, resulting in a correspondingly large donor mass estimate close to what is expected for a main-sequence object filling its Roche-lobe at the orbital period of WZ Sge ($M_{2_{ZAMS}}=0.08M_{\odot}$). 
This is in contrast with previous constraints on the nature of the mass donor that imply a low luminosity object.
Searches for signatures of the mass donor in the red and infrared have so far failed. Indeed, this has been one of the motivations for considering WZ Sge as one of the prime candidates for harboring a degenerate secondary (Howell et al. 2004, Patterson et al. 2001) as the inferred brightness constraints are not consistent with a $0.08M_{\odot}$ main-sequence object. 

An alternative method for estimating the mass ratio of CVs uses the precession periods of the accretion disks as a dynamical probe through the observed superhump periodicities observed in many systems (Patterson et al. 2005b, Knigge 2006). This method has revealed a remarkable correlation between the observed superhump properties and the mass ratio of the binary. Although WZ Sge is considered a calibrator for this relation, with a cited mass ratio of $q=0.05 \pm 0.015$, we point out that we do not have a reliable independent determination of this value. However, it is clear that our implied mass ratio of $\sim 0.09$ if we combine $K_1$=47 km/s with $M_1=0.85M_{\odot}$ is substantially larger than the trend suggested by the superhump behaviour. 

%
It is unclear whether our UV results are significantly affected by a component that distorts the radial velocity curve causing us to over-estimate $K_1$ and thereby over-estimate the donor star mass and the binary mass ratio considerably. Intervening gas that is not in Keplerian rotation with the primary could spoil the diagnostics provided by the UV lines, although the exact driver behind such a component is unclear and the spectrum and its features are quite well described by WD models. Since we know the orientation of the binary fairly well thanks to the stable bright spot eclipses and the detection of the mass donor during outburst, the relative phase offset is significant at the $3-4\sigma$ level.
The inferred donor mass for our favored system parameters ($M_2=0.078 \pm 0.06 M_{\odot}$) corresponds to an $\sim$L2 type star. If we take the Knigge (2006) donor sequence, we find $M_2\sim0.073M_{\odot}$ for a system still evolving towards its period minimum and $M_2\sim0.044M_{\odot}$ for a post period minimum system at the orbital period of WZ Sge. 
Our $M_2$ value is right near the brown dwarf treshold and would suggest that WZ Sge has not bounced back yet from its period minimum. However, an object of that mass should have been detected during previous near-infrared studies of WZ Sge given the expected $M_K\sim10.4$ and the $M_K>12$ constraint from Howell (2004). The absence of any signatures of the mass donor might suggest a massive, but sub-luminous and evolved object instead of an object near its main-sequence.  We could speculate how the low mass donor star may have adjusted to its long history of mass loss, but at this point have no direct evidence for abnormal abundances in the accretion flow for example that might be expected for such an evolved and/or stripped object.
Despite our exploitation of high-resolution HST observations of the glowing white dwarf in WZ Sge, we must still keep an open mind about the system parameters of this important cataclysmic variable. The nature of the observed phase offset and its impact on the absorption line radial velocities need to be clarified. It is clear that additional high resolution UV spectroscopy of the system as it settles into its long quiescent phase could provide useful constraints on the environment near the white dwarf photosphere.

\acknowledgments

DS acknowledges a Smithsonian Astrophysical Observatory Clay Fellowship as well as support through a PPARC/STFC Advanced Fellowship.
The research reported here was based on observations made with the 
NASA/ESA Hubble Space Telescope. Support for this work was provided by NASA through grant number GO-09459.
from the Space Telescope Science Institute. STScI is operated
by the Association of Universities for Research in Astronomy, Inc. under NASA
contract NAS 5-26555.

\end{document}